\documentclass{llncs}
\usepackage{llncsdoc}
\usepackage{epsfig}
\usepackage{graphicx}
\usepackage{hyperref}
\usepackage{bsymb}
\usepackage{alltt}
\usepackage{amsmath}
\usepackage{amssymb}
\usepackage{stmaryrd}
\author{A. Edmunds\inst{1} and J. Colley and M. Butler}
\institute{\textsuperscript{1}University of Southampton, UK\\ \email{ae2@ecs.soton.ac.uk}}
\title{Building on the DEPLOY Legacy: \\Code Generation and Simulation}
\begin{document}
\maketitle
\thispagestyle{empty}
\pagestyle{empty}
%
\begin{abstract}
The RODIN, and DEPLOY projects laid solid foundations for further theoretical, and practical (methodological and tooling) advances with Event-B. Our current interest is the co-simulation of cyber-physical systems using Event-B. Using this approach we aim to simulate various features of the environment separately, in order to exercise deployable code. This paper has two contributions, the first is the extension of the code generation work of DEPLOY, where we add the ability to generate code from Event-B state-machine diagrams. The second describes how we may use code, generated from state-machines, to simulate the environment, and simulate concurrently executing state-machines, in a single task. We show how we can instrument the code to guide the simulation, by controlling the relative rate that non-deterministic transitions are traversed in the simulation. 
\end{abstract}
\section{Introduction}
This paper describes activities undertaken during the early part of the ADVANCE~\cite{advance} project. Building on the RODIN, and DEPLOY projects~\cite{DEPLOY}, we are working to better understand the issues arising in a development when modelling with Event-B, and animating with ProB, in tandem with a multi-simulation strategy. Some of DEPLOY's industrial partners were interested in the formal development of multi-tasking, embedded control systems. We developed an approach for automatically generating code from Event-B models, for these types of systems~\cite{ae2011a}. In this paper we also present an extension to this work.

Event-B uses set-theory, predicate logic and refinement to model discrete systems. The basic structural elements of Event-B models are contexts and machines. Contexts describe the static aspects of a system, using sets, constants, and axioms. The contents of a Context can be made visible to a machine. Machines describe the dynamic aspects of a system, in the form of state variables, and guarded events, which update state. Required properties are specified using the invariants clause. The invariants give rise to proof obligations. 

In the remainder of this section, we describe the Event-B representation of state-machines, and Tasking Event-B, our existing code-generation approach. We introduce a case study in Section~\ref{StopStartModel}. In Section~\ref{CGSM} we describe the new code generation feature for Event-B state-machine diagrams. In Section~\ref{Manip} we describe how we use a single task, generated from state-machines, to simulate the environment and concurrently executing state-machines. We show how we guide the simulation, using additional guards on the transition implementations, to control the relative rate that non-deterministic transitions are traversed. We conclude with Section~\ref{conc}.

\begin{figure}[t]
\centering
\includegraphics[width=0.6\textwidth]{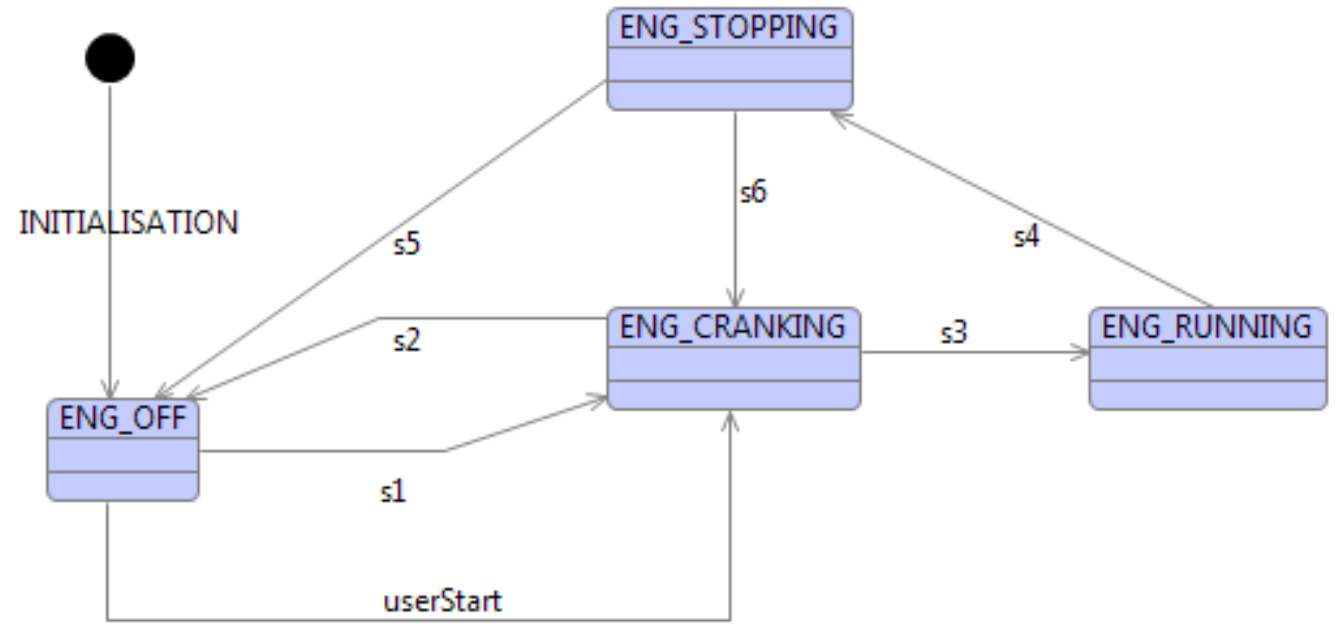}
\caption{EngMode State-machine}
\label{fig:EngSM}
\end{figure}

State-machine diagrams~\cite{EventBSM} can be added to a machine. Each contains an initial state, typically contains one or more transitions, one or more other states, and possibly a final state. A transition 'elaborates' one or more events; that is, a transition describes the atomic state updates that occur during the change from one state to the next. We use an example of an automotive engine stop-start controller, loosely based on~\cite{DEPLOYD38}, to illustrate our approach. The system aims to save fuel by switching the engine off when the car is stationary. Fig.~\ref{fig:EngSM} is an example of a state-machine diagram, \emph{EngMode}. Initially the state-machine is in the \emph{ENG\_OFF} state, and may go the \emph{ENG\_CRANKING} state via transitions \emph{s1} or \emph{userStart}, and so on. In the properties we define `translation type'  as \emph{Enumeration}. The underlying Event-B model, uses a set-partition of the states, as shown below. The current state of the state-machine is recorded in a variable $EngMode \in EngMode\_STATES$, where $EngMode\_STATES$ is a partition of the states of the EngMode state-machine,
\begin{equation}
\begin{split}
\mathit{partition}&(\mathit{EngMode\_STATES, \{ENG\_STOPPING\}},\\
&\mathit{\{ENG\_CRANKING\},\{ENG\_RUNNING\}, \{ENG\_OFF\})}
\end{split}
\label{eq:partition}
\end{equation}
\subsection{Tasking Event-B}
Tasking Event-B~\cite{ae2011a,Edmunds2012a} is an extension to Event-B; where Event-B elements are restricted to implementable types. If required we use decomposition~\cite{decomp2010b,decomp2010c} to separate the system into sub-components. At an appropriate stage we introduce implementation specific constructs to guide code generation. These constructs are underpinned by Event-B operational semantics; Tasking Event-B introduces three main constructs:- AutoTask, Environ, and Shared Machines. AutoTask Machines model controller tasks (in the implementation). Environ Machines model the environment, and Shared Machines provide a protected resource for sharing data between tasks.

\begin{figure}
\centering
\includegraphics[width=0.7\textwidth]{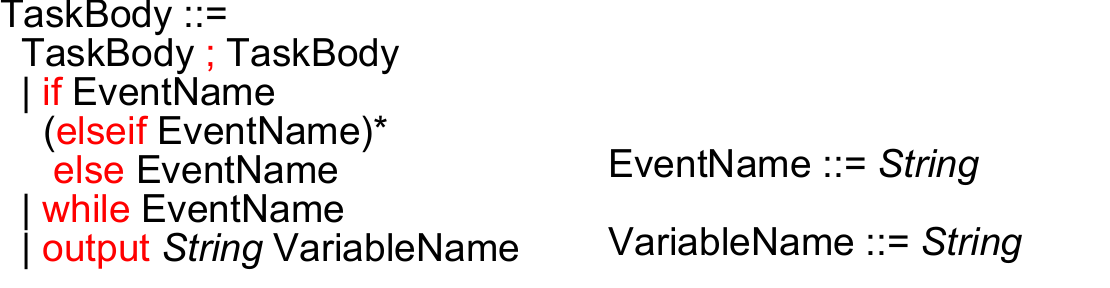}
\caption{Task Body Syntax}
\label{fig:BodySyntax}
\end{figure}

Tasks bodies are specified using the syntax shown in Fig.~\ref{fig:BodySyntax}. We can use (;) sequence , (if-elsif-else) branching, (do) looping, and text output to the console.
\subsection{Translation of a Task Body}
To simplify the discussion, our example uses a single tasking approach. We will not consider here the issue of multi-tasking. We therefore need only to give a brief overview of AutoTask Machine translation, since it will not be synchronized with a Shared Machine. Given an event $E \triangleq g \rightarrow a$, we map action $a$ to a program statement $a'$, and guard $g$ to a condition $g'$, if $g$ exists. The guard should be $\btrue$ for events used in sequences, but may be any implementable predicate for use in branching and looping statements. An example translation of branching follows, where events $e_1 \triangleq g1 \rightarrow a_1$ and $e_2 \triangleq g_2 \rightarrow a_2$, are used in the task body,
\begin{equation}
\begin{split}
&\textbf{if }~e_1~\textbf{else}~ e_2~ \textbf{endif}\\
&\rightsquigarrow\\
&\textbf{if}~ g'_1~ \textbf{then}~ a'_1~ \textbf{else}~ a'_2~ \textbf{end~if};
\end{split}
\notag 
\end{equation}
The branching construct of the task body contains events $e_1$ and $e_2$, and translates to a branching construct in the program code. The guard $g'_2$ does not appear in the code, but to ensure that the modellers intentions are correctly implemented a proof obligation can be generated to ensure that $g_2 \leqv \lnot g_1$. The tool could be augmented to generate proof obligations automatically, to show that branch guards are disjoint and complete.

\section{The Automotive Stop-Start Model}\label{StopStartModel}
\begin{figure}
\centering
\includegraphics[width = 0.6\textwidth]{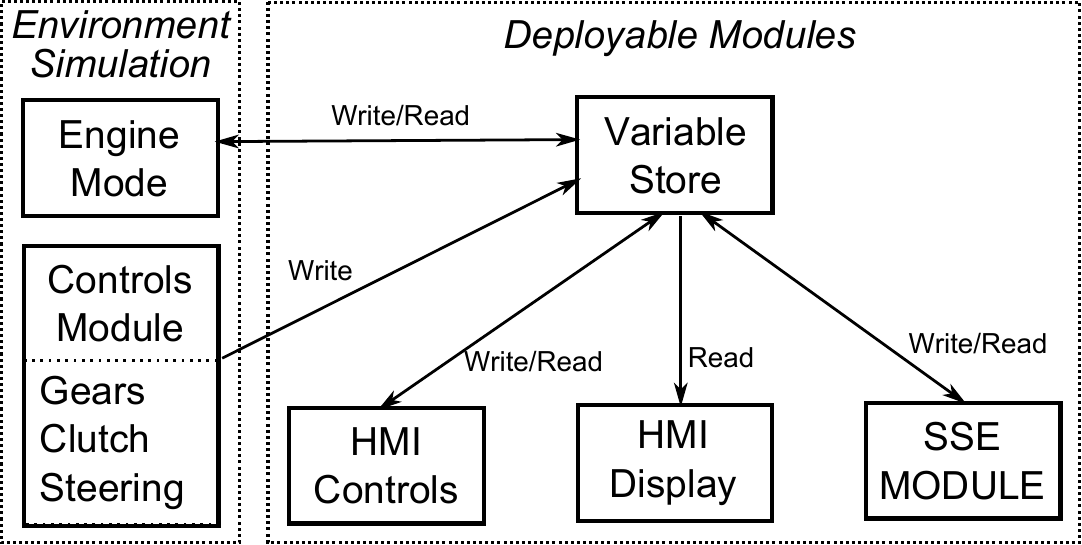}
\caption{Overview of the Stop-Start Architecture}
\label{fig:architecture}
\end{figure}

A typical approach to multi-tasking in hybrid systems, relies on a \emph{write-read-process} protocol. The shared variable store, shown in Fig.~\ref{fig:architecture}, is used by the various modules; to write to, and then read from. In such a system, each task keeps a local copy of the parts of the state that it needs to deal with. In the \emph{write-read-process} protocol, all tasks write to the store, all tasks then read from the store. Only when all tasks have updated their local copies of shared state, can processing take place. The task iterates these steps in a loop. In our tool we simulate the concurrent implementation using \emph{sequential} code generated from a single AutoTask Machine. The deployable modules of Fig.~\ref{fig:architecture} can be implemented in a multi-tasking environment if the execution order of the protocol is preserved.

In our sequential simulation, we use a single AutoTask Machine, which contains both controller and environment state-machines; and define write and read behaviour in the machine's task-body construct. We have already seen the Stop-Start (SSE ) system's \emph{EngMode} state-machine, in Fig.~\ref{fig:EngSM}. In addition to this we have Clutch, Gear and Steering environment state-machines. There are three controller modules, the SSE Module that decides whether to issue stop or start commands based on the engine state, and values determined by the HMI Controls module. HMI Controls monitors the clutch, gear, and steering controls to see if automatic stop or start should be enabled.
\subsection{The Task-body}
We have defined the state-machines of the system and we can now specify the IO between the modules via the shared variable store. The store contains a copy of all of the variables involved in IO between modules. Each of the modules may send data and receive data from the variable store. If we take, as an example, the engine's IO, we output the engine state and speed to the shared variable store. All variables in the store are prefixed `STO\_', and variables in the engine module (other than state names) are prefixed `ENG\_', so the following event updates the shared variable store's copies of the engine state. Each state-machine has a send (\emph{write}) and receive (\emph{read}) event which has the state-machine name and $send$ or $recv$ as a suffix.
\begin{equation}
\begin{split}
\mathit{Eng\_send \triangleq}&\mathit{~STO\_EngMode \bcmeq EngMode} \\
&\pprod \mathit{STO\_EngineSpeed \bcmeq ENG\_EngineSpeed}
\end{split}
\notag
\end{equation}
\subsection{Modelling Starting and Stopping the Engine}
\begin{figure}[t]
\centering
\includegraphics[width = 0.6\textwidth]{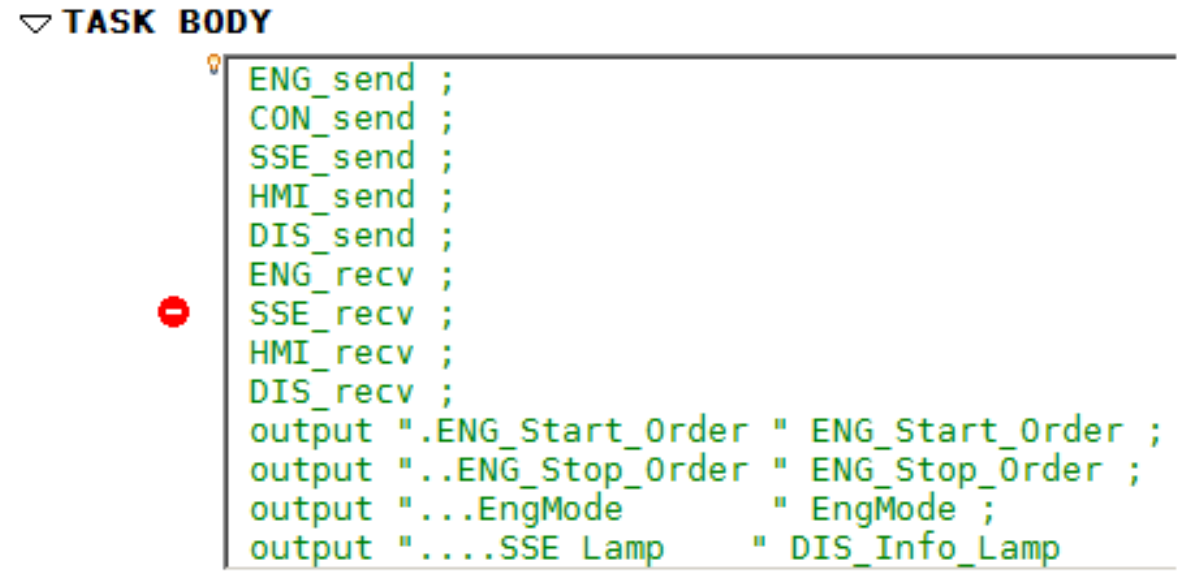}
\caption{The Task Body Specification}
\label{fig:TaskBody}
\end{figure}

 The \emph{EngMode} state-machine keeps track of the engine mode, i.e. off, running, cranking, or stopping. The engine is initially in the \emph{ENG\_OFF} state. We model the ultimate task of the SSE system, the automatic engine start, with the \emph{s1} event. This is enabled after receiving an engine start order from the Stop-Start Controller module (the SSE Module's SSEMode state-machine, introduced later). The \emph{s1} event follows,
\begin{equation}
\begin{split}
\mathit{s1}\triangleq&~\textbf{when}~ \mathit{EngMode = ENG\_OFF \land ENG\_Start\_Order = TRUE} \\
&\textbf{then}~ \mathit{EngMode \bcmeq ENG\_CRANKING} \\
&\textbf{end} 
\end{split}
\notag
\end{equation}
The predicate and action involving \emph{EngMode} are generated automatically in the translation from the state-machine diagram. The guard with \emph{ENG\_Start\_Order} is added by the developer to indicate that the engine should enter the cranking state when a Start Order has been received. The engine may also be started manually, as modelled by the userStart event. When the engine is running at a sufficient rate $s3$ sets the engine state to \emph{ENG\_RUNNING},  
\begin{equation}
\begin{split}
\mathit{s3 \triangleq}&~\textbf{when}~\mathit{ EngMode = ENG\_CRANKING} \\
&\quad\qquad \land \mathit{Eng\_EngineSpeed >= Eng\_Idle\_Speed} \\
&\textbf{then}~\mathit{EngMode \bcmeq ENG\_RUNNING} \\
&\textbf{end} 
\end{split}
\notag
\end{equation}
When the engine is running, it can be stopped automatically by the SSE module. The \emph{HMI\_Controls} module checks to see if it is in neutral gear, steering not-used, and clutch released. If it is, \emph{HMI\_Stop\_EnaT} sets \emph{HMI\_Stop\_Ena} to true. This is eventually passed to the \emph{SSEMode module} via the shared store.
\begin{equation}
\begin{split}
&\mathit{HMI\_Stop\_EnaT \triangleq}~\\
&\textbf{when}~ \mathit{HMI\_Gear = NEUTRAL \land HMI\_Steer = NOT\_USED}\\
&\land \mathit{HMI\_Clutch = RELEASED \land HMI\_ControlsSM = HMI\_OPERATION} \\
&\textbf{then}~ \mathit{HMI\_Stop\_Ena \bcmeq TRUE \pprod HMI\_Strt\_Req \bcmeq FALSE}\\
&\textbf{end} 
\end{split}
\notag
\end{equation}
Event $t7$ elaborates a transition of the SSE state-machine diagram, setting $SSE\_Stop\_Order$ and $ SSE\_Start\_Order$. This is copied to the variable store, and then read by the engine module. 
\begin{equation}
\begin{split}
&\mathit{t7 \triangleq}~\\
&\textbf{when}~ \mathit{SSEMode = SSE\_OPERATION \land SSE\_Stop\_Req = TRUE}\\
&\land \mathit{SSE\_EngMode = ENG\_RUNNING \land SSE\_Stop\_Ena = TRUE}\\
&\textbf{then}~\mathit{SSEMode \bcmeq SSE\_STOPPING \pprod SSE\_Stop\_Order \bcmeq TRUE}\\
& \pprod \mathit{SSE\_Start\_Order \bcmeq FALSE} \\
&\textbf{end} 
\end{split}
\notag
\end{equation}
We specify the sequence of events in the Task Body in the `usual' Tasking Event-B style, seen in Fig.~\ref{fig:TaskBody}. We have specified that send events occur before the read events. This is necessary to ensure the latest state is made available for the state-machine evaluation. The Task Body is periodic, and generates a loop in the implementation. The order of processing is as follows: 1) Initialisation of state. 2) Evaluate state-machines. 3) Send updated values to the variable store. 4) Read updated values from the variable store; then go to 2, and repeat.  The sequence \{4,2,3\}, in the task body, corresponds to the \emph{read-process-write} protocol, which follows initialisation (and initial sends to the variable store). Fig.~\ref{fig:TaskBody} also shows the \emph{output} clause, for text output to the console. 
The next section provides details of the translation to Ada code.
\section{Translating State-Machines to Ada Code}\label{CGSM}
To illustrate the translation process we  show the Ada implementation, we have seen how state-machine states are modelled by an enumeration partition, and we use this in the implementation. The partition of Equation~\ref{eq:partition} is translated to the following Ada code.
\begin{equation}
\begin{split}
&\textbf{package}~ \mathit{StopStart01b\_Globals}~ \textbf{is} \\
&\textbf{type}~ \mathit{EngMode\_STATES}~ \textbf{is} (\\
&\qquad \mathit{ENG\_STOPPING,~ ENG\_CRANKING},\\
&\qquad \mathit{ENG\_RUNNING,~ ENG\_OFF)};\ldots
\end{split}
\notag
\end{equation}
We create the package \emph{StopStart01b\_Globals} to store the global constants and types. The type \emph{EngMode\_STATES} is an enumeration of the state-machine states. Recall also, that we generate a state variable $EngMode$ which is typed as $EngMode \in EngMode\_STATES$, to keep track of the state; it has the initial value \emph{Eng\_OFF}. We use the diagram and the initialisation event to generate the following code:
\begin{equation}
\mathit{EngMode : EngMode\_STATES \bcmeq ENG\_OFF;}
\notag
\end{equation}
The main program invokes the state-machine implementations in a loop, once per cycle. Each state-machine diagram maps to a procedure. State-machine procedures are called exactly once before the sends to, and reads, from the variable store. The evaluation of each state-machine procedure is independent of the other state-machines, since each keeps a local copy of the state, copied from the variable store. Each state-machine procedure has a state variable $v$, states $w_i$, and implemented actions $a_i$. To each state-machine procedure, we add to a \emph{case} statement,
\begin{equation}
\begin{split}
&\textbf{case}~ v~ \textbf{is} ~\textbf{when}~ w_1 => a_1;\\
&\textbf{when}~ w_2 => a_2; ~ \ldots\\
&\textbf{when}~ w_n => null;\\
\end{split}
\notag
\end{equation}
 Translation of our example gives rise to the following code,  
\begin{equation}
\begin{split}
&     \textbf{procedure}~ \mathit{EngModestateMachine}~ \textbf{is}\\
&      \textbf{begin}\\
&      \textbf{case}~ \mathit{EngMode}~ \textbf{is}\\
& \quad     \textbf{when}~ \mathit{ENG\_STOPPING} =>\\
& \quad           \textbf{if}~ \mathit{((ENG\_EngineSpeed = 0))}~ \textbf{then}\\
& \qquad             	\mathit{EngMode~ :=~ ENG\_OFF; \qquad  -- ~s5}\\
& \quad           \textbf{elsif}~ \mathit{((ENG\_Start\_Order = true))}~ \textbf{then}\\
& \qquad              \mathit{EngMode~ :=~ ENG\_CRANKING; ~~-- ~s6}\\
&\quad            \textbf{else}~ null;\\
&\quad            \textbf{end~ if};\\
&\quad         \textbf{when}~ \mathit{ENG\_CRANKING} =>\\
&\quad            \textbf{if}~ \mathit{((ENG\_EngineSpeed = 0))}~ \textbf{then}\\
&\qquad               \mathit{EngMode~ :=~ ENG\_OFF; \qquad -- ~s2}\\
&\quad            \textbf{elsif}~ \mathit{((ENG\_EngineSpeed >= Eng\_Idle\_Speed))}~ \textbf{then}\\
&\qquad               \mathit{EngMode~ :=~ ENG\_RUNNING;  ~~--~ s3}\\
&\quad            \textbf{else}~ null;\\
&\quad            \textbf{end~ if};\\
&\quad         \textbf{when}~\mathit{ENG\_RUNNING} => \ldots\\
&         \textbf{end~ case};\\
&      \textbf{end}~ \mathit{EngModestateMachine};
\end{split}
\notag
\end{equation}
We can see that each of the case's $when$ statements contains a branching statement. This is because each state of the state-machine has at least two branches; a do-nothing transition, plus one or more outgoing transitions. The do-nothing transition is not explicitly shown on the diagram. A do-nothing transition can be added to each state, since adding a \emph{skip} event is a valid refinement. It is implemented by the \textbf{else null;} branch. Other branches are translated from states with more than one outgoing transition. This may be seen in the \emph{ENG\_STOPPING} case in the example. The  branch conditions are mapped from the guards of the events ($s5$ and $s6$) that elaborate the outgoing transitions. 
\section{Manipulating State Machine Transitions}\label{Manip}
The generated code from our example is compiled to an executable file and run. We have implementable code for the controller state-machines, and a simulation of the environment from the environment state-machines. When executing, we find that most of the state remains unexplored, and this is due to the non-determinism in the state-machines. This section identifies how we can guide a simulation, by reducing the non-determinism in the generated state-machine by modifying the branch conditions.

For the controller state-machines, each state's outgoing transitions are disjoint and complete; in other words, a transition is always taken in the simulation. However, in the environment, it is unlikely that the clutch changes state so frequently. We do have the implicit \emph{do-nothing} transition on environment state-machine states, but we need this to happen more often than the other transitions. We must have some control over the relative rate that non-deterministic transitions are traversed in the simulation. As it stands, any outgoing transition is equally likely to occur. To solve this in the simulation, we introduce an enabling variable $q\in 0\upto n$ and a random variable $r \in 0\upto n$, and use the random variable in a case-statement's branch conditions. Variable $q$ is calculated once at the beginning of the simulation, but a new random variable $r$ is calculated at each state-machine evaluation. The event $g \mapsto a$ in Event-B terms is implemented as a branch $g \land r=q \mapsto a$ in a case-statement. 

We now suggest how we may generate, and use the variables $q$ and $r$ in simulation. This aspect is work in progress, but we believe the approach will be useful for generating test scenarios, and therefore will help to improve test coverage. By adding a guard to the branch condition we can influence the path taken through the code during simulation. In effect, we reduce the non-determinism in the state-machines, which allows us to guide the simulation, and therefore the exploration of the state-space. 

\begin{figure}[t]
\centering
\includegraphics[width=0.8\textwidth]{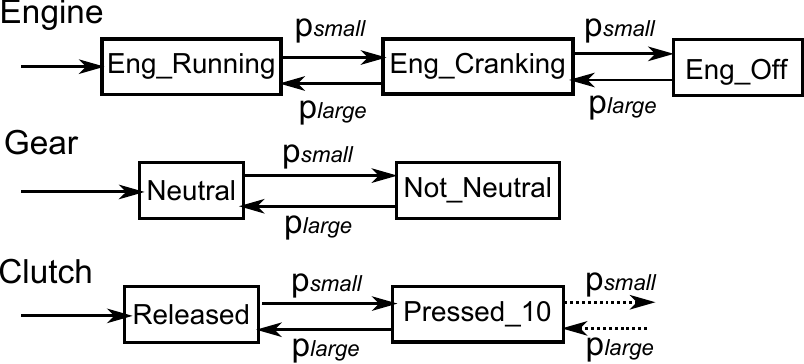}
\caption{Controlling the Simulation}
\label{fig:probs}
\end{figure}

One question is, how to choose a value of $n$? We could base it on the total number of outgoing transitions of the state involved, but this would not give a large enough value. A typical state may have four transitions, plus a \emph{do-nothing} transition, so a random number $r \in 0\upto 4$ could be used. However, we wish to manipulate the probability of a branch being taken, so that a branch is very unlikely to be taken; therefore, a much larger value for $n$ is required. So, we calculate $n$ based on the number of tests that would be required, for test coverage of all transitions, in all states. Likewise, the value of $q$ must be unique within the case-statement; we just allocate an arbitrary, but unique value, close to $n$. In future work we will investigate how we could modify $n$ during simulation runs, and use this value to reduce the probability of a simulation traversing previously explored state.
In the code fragment below, we add the probabilistic condition to the branch of the case-statement, where $r = StartStop01b\_random$ ($StartStop01b\_random$ is a random variable in the implementation code) and $q = 3990$.
\begin{equation}
\begin{split}
&	\textbf{case}~ \mathit{EngMode}~ \textbf{is}  \\
&	\textbf{when} ~\mathit{ENG\_STOPPING} => \\
&	\textbf{if}~ \mathit{((ENG\_EngineSpeed = 0))~ and~ (StopStart01b\_random = 3990)}~ \textbf{then} \\
&\qquad		\mathit{EngMode := ENG\_OFF;} \ldots 
\end{split}
\notag
\end{equation}
Adding the branch condition gives us control over the likelihood that a particular transition from a state will be taken when the state-machine is evaluated. We manually modify the conditions, to affect the behaviour of the simulation. We may wish to focus on exploring the state in a particular region. For instance, to test an engine-stop scenario, we require that the engine is in the ENG\_ RUNNING state, the gear is in NEUTRAL, the clutch is in the RELEASED state, and the steering NOT\_USED. Fig.~\ref{fig:probs} shows that we want large probabilities of transitions leading to the states that we want, and small ones departing.

For a given simulation run we can define $attracting$ and $repelling$ states. Here, ENG\_RUNNING is an attracting state; that is, we want the state-machine to be in that state or moving towards it most of the time. To achieve this we can adjust the branch conditions, to increase the probability of the transitions that lead to that state, being taken. For instance to increase the probability of the engine going from ENG\_OFF to ENG\_CRANKING we can modify the statement to read (StopStart01b\_random $<= $ 3990). In addition to this, we propose to record the navigated transitions, for transition coverage analysis. So, we will be able to use the data also, to guide the simulation. We show two simulation runs here, with the text output defined in the Task Body, \emph{Run1} uses the `unmodified', generated code; it simply loops and never reaches the ENG\_RUNNING state. With the branch conditions modified, as described, \emph{Run 2} shows the simulation cycling from $ENG\_RUNNING$ to $ENG\_OFF$; and with the indicator lamp changing to inform the driver of the situation.

\begin{minipage}{0.4\linewidth}
\begin{tabular}{|l}
Run 1 \\
\hline
\rule{0pt}{3ex}
.ENG\_Start\_Order  FALSE \\
..ENG\_Stop\_Order  FALSE \\
...EngMode        ENG\_OFF \\
....SSE Lamp     OFF \ldots
\end{tabular}
\end{minipage}
\begin{minipage}{0.5\linewidth}
\begin{tabular}{|l}
Run 2 \\
\hline
\rule{0pt}{3ex}
.ENG\_Start\_Order  FALSE\\
..ENG\_Stop\_Order  TRUE\\
...EngMode        ENG\_RUNNING\\
....SSE Lamp     OFF\\
.ENG\_Start\_Order  FALSE\\
..ENG\_Stop\_Order  TRUE\\
...EngMode        ENG\_STOPPING\\
....SSE Lamp     ORANGE\_STOP\\
.ENG\_Start\_Order  FALSE\\
..ENG\_Stop\_Order  TRUE\\
...EngMode        ENG\_OFF \ldots
\end{tabular}
\end{minipage}
\section{Conclusions}\label{conc}
We have shown how we generate Ada code from State-machines, and illustrated the approach with a case study based on an automotive engine controller, automatic stop-start system. We describe how we simulate the environment, and a multi-tasking implementation. We gain an insight into how we adjust the conditions to provide meaningful simulation runs, which should be useful in the ADVANCE project. In future work we intend to record the transition coverage, and feed this back to the simulator, to ensure all transitions are covered. We will also investigate the interaction between the generated code, environment simulations, and ProB.
%
%
\label{sect:bib}
\bibliographystyle{plain}
\bibliography{MyBibTex}


\end{document}